\documentclass[prl,twocolumn,aps,groupedaddress,showpacs]{revtex4}

\usepackage{epsfig, amssymb}

\def\be{\begin{eqnarray}}
\def\ee{\end{eqnarray}}
\def\bee{\begin{eqnarray*}}
\def\eee{\end{eqnarray*}}

\begin{document}

\title{Completeness of the classical 2D Ising model and universal quantum computation}

\author{M. Van den Nest$^1$, W. D\"ur$^{1,2}$, and H. J. Briegel$^{1,2}$ }

\affiliation{$^1$ Institut f\"ur Quantenoptik und Quanteninformation der \"Osterreichischen Akademie der Wissenschaften, Innsbruck, Austria\\
$^2$ Institut f{\"u}r Theoretische Physik, Universit{\"a}t
Innsbruck, Technikerstra{\ss}e 25, A-6020 Innsbruck,
Austria}
\date{\today}

\date{\today}
\def\makeheadbox{}

\begin{abstract}
We
prove that the 2D Ising model is \emph{complete} in the sense that the
partition function of any classical $q$-state spin
model (on an arbitrary graph) can be expressed as a special instance of the
partition function of a 2D Ising model with complex inhomogeneous
couplings and external fields. In the case where the
original model is an Ising or Potts-type
model, we find that the corresponding 2D square lattice
requires only polynomially more spins w.r.t the original
one, and we give a constructive method to map such models to the 2D Ising model. For more general models the overhead in system size may be exponential. The results are
established by connecting classical spin models with
measurement-based quantum computation and invoking the
universality of the 2D cluster states.
\end{abstract}


\maketitle

{\it 1.--- Introduction.}
Classical spin models such as the Ising and Potts models
are widely studied in statistical physics, as they provide
important toy models for magnetism and as they can be
mapped to numerous interesting problems in physics and
mathematics \cite{Wu83,Soxx}. The geometry of a model, in particular its
spatial dimension, plays an important role with respect to
the physical properties of the system and the possibility
of finding (approximate) solutions. For instance, it is
known that evaluation of the partition function of the
Ising model with magnetic fields  is easy in 1D, while on a
2D square lattice this problem is already NP-hard
\cite{Ba82}.

In this paper we study the interrelations between classical
$q$-state spin models on different geometries (or graphs),
and find that the 2D Ising model (which has $q=2$) plays a
distinguished role in this study.
We consider mappings that leave the partition function---and hence all
thermodynamical quantities, such as free energy or magnetization, derived from it---
invariant (see also \cite{QIT_spin, Va07}).
As the main result of this paper, we
prove that the 2D
Ising model is \emph{complete} in the sense that the
partition function of {\em any} classical $q$-state spin
model can be expressed as a special instance of the
partition function of a 2D Ising model with inhomogeneous
couplings.
More precisely, given a partition function $Z_G$ of a
$q$-state spin model on an arbitrary graph---which may be,
e.g., a lattice of arbitrary dimension or involve
long-range interactions---there exists a 2D square lattice
of enlarged size, and suitably tuned nearest-neighbor
coupling strengths and magnetic fields, such that the
partition function of the Ising model on this lattice
specializes to $Z_G$. Furthermore, in the case where the
original model on the graph $G$ is an Ising or Potts-type
model, we find that the corresponding 2D square lattice
requires only polynomially more spins w.r.t the original
one. For more general models the overhead in system size
may be exponential.
However, one important remark needs to
be made: in order to achieve this
result, one has to allow for \emph{complex} couplings in
the 2D partition function---thus leaving the ``physical''
regime of the model.

The results are proven by relating the problem at hand to
insights from quantum information theory, more particularly
to the area of \emph{measurement-based quantum computation}
(MQC). The latter is a recently established paradigm for
quantum computation where quantum information is processed
by performing sequences of single-qubit measurements on a
highly entangled resource state \cite{Ra01}. In order to obtain our results, we first prove
that the Ising partition function on an arbitrary graph
(with external field) can be written as the overlap between
an entangled quantum state and a complete product
state---thus generalizing a construction which we
introduced in Ref. \cite{Va07}; see also Ref. \cite{Br06}.
This formulation allows us to make a connection with MQC.
In particular, we prove that the entangled state
corresponding to the Ising model on a 2D square lattice, is
(a variant of) the \emph{2D cluster state} \cite{Ra01b}.
The latter is known to be a \emph{universal} resource state
for MQC in the sense that every quantum state can be
obtained by performing a suitable sequence of single-qubit
measurements on a sufficiently large 2D cluster state. This
quantum universality feature of the 2D cluster states leads
to the result that the 2D Ising model is complete in the
sense specified above.

{\it 2. Classical Ising model.---}
We consider the classical Ising model involving $N$ two-state spins
$(s_1,s_2\ldots s_N)\equiv {\bf s}$, where $s_a = \pm 1$. The spins
interact pairwise according to an interaction pattern specified by a
graph $G=(V,E)$ with vertex set $V$ and edge set $E$, and the
coupling strengths are denoted by $J_{ab}$. Moreover, the spins are
subjected to local magnetic field terms $h_a$. The Hamiltonian of
the system is given by $H_G({\bf s}):=-\sum _{\{a, b\} \in E} J_{ab}
s_{a}s_{b} - \sum_{a \in V} h_a s_a$. In other words, we consider a
general inhomogeneous Ising model on an arbitrary graph. The
partition function $Z_G$ is defined by $ Z_{G}(\{J_{ab},h_a\}) :=
\sum_{\bm s} e^{-\beta H_G({\bm s})}, $ where $\beta=(k_BT)^{-1}$,
with $k_B$ the Bolzmann constant and $T$ the temperature.

\begin{figure}[ht]
\begin{center}
\begin{picture}(210,58)
\put(0,-5){\includegraphics[width=0.75\columnwidth]{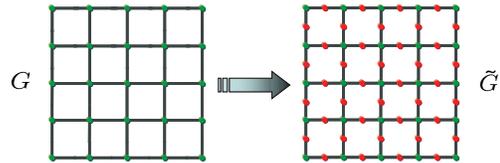}}
\end{picture}
\caption {\label{Fig1} (Color online) Decorated graph $\tilde G$ (right)
corresponding to a 2D lattice $G$ (left).  Green (dark) dots indicate
vertices
originating from the vertices $V$ of $G$, while red (light) dots
indicate vertices $V_E$ originating from edges $E$ of $G$.}
\end{center}
\end{figure}

{\it 3. Quantum formulation.---}
We now show how the partition function $Z_G$ can be expressed in a
quantum physics language. Let $\tilde G$ be the graph with
$n=|V|+|E|$ vertices and $2|E|$ edges which is obtained from $G$ by
adding at each edge $\{a, b\}\in E$ an additional vertex $ab$ and
thus ``splitting every edge in half'' (see Fig. \ref{Fig1}). We will
call $\tilde G$ the decorated version of $G$. The vertex set of
$\tilde G$ is thus given by the union of the original vertex set $V$
of $G$ and the set $V_E = \{ ab\ |\ \{a, b\}\in E\}$ corresponding
to edges of $G$---note that we label the vertices in $V_E$ by double
indices, indicating their origin in the corresponding edge of $G$.
We now consider an $n$-qubit state $|\varphi_{\tilde G}\rangle$,
defined on a set of qubits labelled by $V\cup V_E$, which is defined
to be the graph state \cite{He06, noteGraphstate} associated with
the decorated graph $\tilde G$. In particular, $|\varphi_{\tilde
G}\rangle$ is the (unique) joint fixed point of the $|V| + |E|$
stabilizing operators $K_{a}$ and $K_{ab}$, \be\label{K}
K_a&=&X^{(a)} \prod_{b:\{a,b\}\in E} X^{(ab)} \nonumber\\ K_{ab}&=&
Z^{(ab)} Z^{(a)} Z^{(b)}, \ee $\mbox{for every } a \in V$, and
$\mbox { for every } e=\{a,b\}\in E$. Here $X$ and $Z$ denote the
Pauli spin matrices, and the upper indices indicate on which qubit
is acted.

We can now express the partition function as follows: \be
\label{overlap} Z_{G}(\{J_{ab},h_a\}) = 2^{|V|/2}\ \cdot
\langle \alpha|\varphi_{\tilde G}\rangle. \ee In this
expression, \be |\alpha\rangle =\left(\bigotimes_{ab\in
V_E}|\alpha_{ab}\rangle \bigotimes_{a\in
V}|\alpha_a\rangle\right) \ee is a complete product state
specifying the coupling strengths of the Ising model. In
particular,  $|\alpha_{ab}\rangle = e^{\beta
J_{ab}}|0\rangle + e^{-\beta J_{ab}}|1\rangle$ is an
(unnormalized) one-qubit state (acting on qubit $ab$)
determined by the interaction strength between particles
$a$ and $b$. Similarly, $|\alpha_{a}\rangle = e^{\beta
h_a}|0\rangle + e^{-\beta h_a}|1\rangle$ is an
(unnormalized) one-qubit state (acting on qubit $a$)
determined by the local magnetic field at particle $a$.
Expression (\ref{overlap}) shows that $Z_G$ can be obtained
by calculating the inner product of the graph state
$|\varphi_{\tilde G}\rangle$ and a complete product state.
The choice of the product state allows one to specify the
couplings of the Hamiltonian and the temperature, while the
structure of the graph state reflects the interaction
pattern.

To show that Eq. (\ref{overlap}) holds, we use that
$|\varphi_{\tilde G}\rangle$ can be written as $|\varphi_{\tilde
G}\rangle\propto \sum_{\bf t} |B^T{\bm t}\rangle|\bm t\rangle$,
where ${\bm t}$ is a binary vector of length $|V|$, $B$ is the
incidence matrix of the graph $G$, and by writing out the sum
$\langle \alpha|\varphi_{\tilde G}\rangle$. The construction of
$|\varphi_{\tilde G}\rangle$ can be viewed as a generalization of
the one we introduced in Ref. \cite{Va07}. While in Ref. \cite{Va07}
each qubit was associated with an edge of the graph $G$, here we
have two types of vertices: one subset $V_E$ associated to edges
(``edge-qubits'') and one to vertices $V$ (``vertex-qubits''). This
enlarging of the system size allows one to treat also local terms in
the Hamiltonian (whereas Ref. \cite{Va07} only dealt with zero
external field). In addition, the stabilizer of the state
$|\varphi_{\tilde G}\rangle$ can be immediately obtained from the
graph $G$ describing the interaction pattern (or its decorated
version $\tilde G$), as in Eq. (\ref{K}) and Fig. \ref{Fig1}.

{\it 4. MQC and the 2D cluster states.---}
We now turn our attention to measurement-based (or: ``one-way'') quantum computation, and establish a relation to the partition function of the 2D classical Ising model via Eq. (\ref{overlap}).

The one-way quantum computer \cite{Ra01} is a recently
developed model for quantum computation, where computations
are realized by performing single-qubit measurements on a
highly entangled substrate state called the 2D cluster
state $|{\cal C}\rangle$ \cite{Ra01b}; the latter is a
graph state \cite{He06} associated to a 2D square lattice
${\cal C}$.

A particular feature of the one-way
quantum computer is that it is {\emph universal}. This means that any $n$-qubit quantum state can be
prepared, up to local unitary Pauli operations, by
performing sequences of single-qubit measurements on a
$d\times d$ cluster state $|{\cal C}\rangle$ of
sufficiently large system size $M=d^2$.
This property of the 2D cluster states immediately implies
that every $n$-qubit quantum state $|\psi\rangle$ can be
written in the following way:
\be
\Sigma|\psi\rangle =2^{(M-n)/2} \left(I\otimes \langle\beta|\right)|{\cal
C}\rangle. \label{psifromcluster}
\ee
This formula
represents one ``measurement branch'' of a one-way
computation performed on an $M$-qubit cluster state,
yielding the state $|\psi\rangle$ (up to a local operation
$\Sigma$) as an output state on the subset of qubits which
has not been measured. The dual product state
$\langle\beta|=\bigotimes_j\langle\beta_j|$, which acts
only on the measured qubits, is determined by the bases and
the outcomes of the different steps in the computation. The
local unitary operator $\Sigma$ (``correction
operator'') acts on the unmeasured qubits (i.e., on the
Hilbert space of $|\psi\rangle$); the tensor factors of
$\Sigma$ are always instances of Pauli operators:
$\Sigma_i\in\{I, X, Y, Z\}$. The prefactor $2^{(M-n)/2}$
reflects the fact that the success probability of every
measurement branch is $2^{n-M}$.

As proved in Ref. \cite{Ra01}, for all $n$-qubit states
$|\psi\rangle$ that can be efficiently prepared in the
circuit model, i.e., by a polynomial sequence
of  two-qubit gates, the required size $M$ of the cluster state
in Eq. (\ref{psifromcluster}) scales polynomially with the
number of qubits: $M \propto {\rm poly}(n)$. Moreover, in
this case the measurement bases $|\beta_j\rangle$ as well
as the correction operations $\Sigma$ can be efficiently
determined.
Since any graph state on $n$ qubits can be prepared using
at most $O(n^2)$  controlled-phase gates \cite{He06}, it
follows that an arbitrary $n$-qubit graph state
\cite{noteGraphstate} can be written in the form
(\ref{psifromcluster}) with $M=\mbox{ poly}(n)$.
Furthermore, for the preparation of graph states every
single-qubit state $|\beta_j\rangle$ can always be chosen
to be one of the $X$-, $Y$- and $Z$-eigenstates.

Also $|\varphi_{\tilde{\cal C}}\rangle$ (i.e.,
the state $|\varphi_{\tilde G}\rangle$ where $G\equiv {\cal
C}$ is the 2D square lattice) is a universal resource. This
is because the 2D-cluster state $|{\cal C}\rangle$ can be
deterministically generated from $|\varphi_{\tilde{\cal
C}}\rangle$ (up to a local correction) by performing
single-qubit $Y$-measurements on all qubits in $V_E$. This
fact was already noted in \cite{Br06}. As a consequence,
one has \be\label{universality_decorated} \Sigma'|{\cal
C}\rangle = 2^{|E|/2}\left(I\otimes \langle
0_Y|^{V_E}\right) |\varphi_{\tilde{\cal C}}\rangle,\ee
where $|0_Y\rangle^{V_E}$ is a tensor product of the $(+1)$-eigenstate of $Y$ on all edge-qubits, and
$\Sigma'$ is a local correction.

{\it 5. Universality of the 2D Ising model.---}
We are now ready to establish the connection  between the
evaluation of Ising partition functions and universal MQC.
To this aim, consider the Ising model on a graph $G$. The
partition function $Z_G$ can be expressed in the form
(\ref{overlap}). Now consider the following procedure.

First, the  graph state $|\varphi_{\tilde G}\rangle$ is
written in the form  (\ref{psifromcluster}) when taking
$|\psi\rangle\equiv |\varphi_{\tilde G}\rangle$. Together
with Eq. (\ref{universality_decorated}), this implies that
the partition function $Z_G$ can be written as \be
Z_G(\{J_{ab}, h_a\})= A \cdot \langle \gamma|\varphi_{\tilde{\cal
C}}\rangle,\label{Z_G2D} \ee where $A$ is a constant and
$|\gamma\rangle$ is a product state,
$
|\gamma\rangle= \Sigma|\alpha\rangle \otimes
\Sigma'|\beta\rangle\otimes |0_Y\rangle^{V_E}. $ Note
that, as $|\varphi_{\tilde G}\rangle$ is a graph state, the
system size of the 2D cluster state grows polynomially with
the size of $G$. Furthermore, $|\beta\rangle$ consists of
$X$-, $Y$- and $Z$-eigenstates.

Now, applying Eq. (\ref{overlap}) to the 2D Ising model,
the overlap between $|\varphi_{\tilde{\cal C}}\rangle$
 and a complete product state corresponds to a 2D Ising partition function $Z_{2D}$, evaluated in certain couplings $\{J_{ij}', h_i'\}$ determined by $|\gamma\rangle$. %
This allows us to conclude that $Z_G$ can be written as
follows;
\be\label{universality_ising} Z_G(\{J_{ab}, h_a\})\propto
Z_{\mbox{\scriptsize{2D}}}(\{J_{ij}', h_i'\}).\ee
In other words, \emph{the Ising partition function on an
arbitrary graph can be recovered as a special instance of
the Ising partition function on a 2D square lattice}.

Note that, in the above sequence of arguments, one step is particularly crucial, namely the universality of the 2D cluster states: this property is used to ``map'' an \emph{arbitrary} state  $|\varphi_{\tilde G}\rangle$, and hence the associated partition function, to the 2D cluster state, i.e., all states can be ``reduced'' to this single structure.

We give a few remarks regarding this construction. In Eq.
(\ref{Z_G2D}), note that the product state $|\gamma\rangle$
is determined by both the interaction graph $G$ and the
couplings $\{J_{ab}, h_a\}$ of the original model. For, on
the one hand, it contains the states $|\alpha_{ab}\rangle$
and $|\alpha_a\rangle$ encoding the couplings of the
original model; on the other hand, $|\gamma\rangle$
contains states $|\beta_j\rangle$ and $|0_Y\rangle$
corresponding to the sequence of one-qubit measurements
which are to be implemented in order to generate
$|\varphi_{\tilde G}\rangle$ from the universal resource
$|\varphi_{\tilde{\cal C}}\rangle$. In going from Eq.
(\ref{Z_G2D}) to Eq. (\ref{universality_ising}), the state
$|\gamma\rangle$ in turn determines the couplings in which
the 2D model is to be evaluated. Note that the decorated
cluster state $|\varphi_{\tilde{\cal C}}\rangle$ has
vertex-qubits and edge-qubits. The factors of
$|\gamma\rangle$ acting on the edge-qubits determine the
pairwise interactions $J_{ij}'$, whereas the factors of
$|\gamma\rangle$ acting on the vertex-qubits determine the
external fields $h_i'$. The tensor factors of
$|\gamma\rangle$ which act on the edge-qubits are all equal
to $|0_Y\rangle \propto |0\rangle + i|1\rangle$. This
implies in particular that, in (\ref{universality_ising}),
only \emph{homogeneous} pairwise couplings $J_{ij}'$ need
to be considered. Furthermore, due to the imaginary unit
``$i$'' in $|0_Y\rangle$, these couplings generally lie in
a \emph{complex} parameter regime; in particular, one can
show that $\beta J'_{ij}= -i \pi/4$ is a correct choice.
Also, the fact that the $J_{ij}'$ can be chosen to be
homogeneous implies that all information regarding the
pairwise couplings $J_{ab}$ and external fields $h_a$ of
the original model, \emph{and} the graph $G$ of this model,
will be encoded in the factors of $|\gamma\rangle$ acting
on the vertex-qubits, and thus in the external fields
$h_i'$ (which will typically be inhomogeneous).
We further remark that the part  of $|\gamma\rangle$ acting
on the vertex-qubits generically also corresponds to
complex interaction strengths $h_i'$ (e.g., $|\beta\rangle$
may contain $Y$-eigenstates). A special role is played by
those factors of $|\beta\rangle$ which are equal to
$Z$-eigenstate $|0\rangle\propto e^{\infty}|0\rangle + e^{-\infty}|1\rangle$.
These states give rise to ``infinitely large'' external fields at the
corresponding vertices, which effectively corresponds to a
boundary condition.

In conclusion, the  universality of the 2D cluster states
$|\varphi_{\tilde{\cal C}}\rangle$ in the context of MQC,
implies that the Ising partition function  on {\em any}
graph can be expressed as a special instance of a
(polynomially enlarged) 2D Ising model with complex,
homogenous pairwise interactions and complex, inhomogenous
external fields. Note that even though such complex
interaction strengths do not correspond to physical models,
considering the partition function as a function with
complex arguments is commonly done, e.g., in the context of
evaluating the Tutte polynomial or finding
(complex) zeros of $Z_G$ to identify phase transition
points \cite{Soxx}.

{\it 6. Generalizations to $q$-state models.---}
Our results can also be generalized to $q$-state spin
models such as the Potts model \cite{Wu83}. We showed in
Ref. \cite{Va07} that the partition function of a $q$-state
Potts model on a graph $G=(V, E)$ can be written as the
overlap between a stabilizer state $|\varphi_{\tilde
G}^q\rangle$ and a complete product state $|\chi\rangle$:
$Z_G\propto \langle \chi |\varphi_{\tilde G}^q\rangle$.
Similar to the treatment of the Ising model, the state
$|\varphi_{\tilde G}^q\rangle$ depends only on the graph,
and the state
$|\chi\rangle=\bigotimes_{ab}|\chi_{ab}\rangle$
is a complete product state depending only on the couplings
of the model. However, the main difference is that the
single-particle systems are no longer qubits, but
$q$-dimensional systems.
E.g., one finds $|\chi_{ab}\rangle = e^{\beta
J_{ab}}|0\rangle + \sum_{k=1}^{q-1}|k\rangle$ \cite{Va07}.
Interestingly, the partition function of such a $q$-state
model (for arbitrary graphs) can again be expressed as a
special instance of the partition function of the 2D-Ising
model (with $q$ = 2) and complex parameters---again using
the connection to MQC. To achieve this this, we use that
any $q$-dimensional product state can be mapped by a
suitable unitary operation to a product state of
$m_q=\lceil\log_2 q\rceil$ qubits; e.g.,
$|\chi_{ab}\rangle=U_{ab}^\dagger|0\rangle^{\otimes m_q}$. As $q$
is fixed, the unitary $U_{ab}$ can be prepared with a
constant number of two-qubit gates. Being a stabilizer
state, $|\varphi_{\tilde G}^q\rangle$ is preparable by a
poly-sized (qubit) circuit.
It follows that $Z_G$ can be written as
the inner product of an efficiently preparable state
$|\varphi\rangle := \bigotimes_{ab} U_{ab}
|\varphi_{\tilde G}^q\rangle$ (which is now regarded as a
multi-qubit state) with a product state $|0\rangle^{\otimes m_q |E|}$.
The universality of $|\varphi_{\tilde {{\cal C}}}\rangle$
for MQC now implies that $|\varphi\rangle$ can be obtained
by performing single-qubit measurements on a polynomially
enlarged cluster state $|\varphi_{\tilde {\cal C}}\rangle$.
In particular, Eq. (\ref{psifromcluster}) can be applied to
$|\psi\rangle \equiv|\varphi\rangle$. Using a similar
argument to Section 5, this implies that the Potts model
partition function is a special instance of the
partition function of a polynomially enlarged 2D-Ising
model with properly tuned complex parameters and two-state
spins.

The above strategy can even be applied to $q$-state models beyond
the Potts model, e.g. to all models on directed graphs where the
Hamiltonians are arbitrary functions of the difference (modulo $q$)
between spin values, including arbitrary local terms, while still
obtaining a 2D-Ising model with polynomially more spins. Even more
generally, one can verify that the partition function of an {\em
arbitrary} $q$-state spin model (with finite $q$), where arbitrary
pairwise or even $k$-body interactions with bounded $k$ are allowed,
can be written as the overlap between a suitable quantum state and a
product state. This immediately implies that every partition
function can be expressed as a special instance of the 2D-Ising
model. However, in general an exponential overhead may be required.

We further remark  that the 2D square lattice does {\em not} play a
special role in this context: there are many other models with a
similar completeness property \footnote{This issue is similar to the
observation that 3-SAT does not play a distinguished role in the set
of NP-complete problems, as there are infinitely many NP-complete
problems. In this sense, 3-SAT is not distinguished, but in
nevertheless often serves the role of a standard.}. For example, all
Ising models on a  graph $G$ whose associated graph state
$|\varphi_{\tilde G}\rangle$ is a universal resource for MQC, allows
one to draw the same conclusions as for the 2D square lattice.
Examples of such other universal models for MQC include e.g.
hexagonal, triangular and Kagome lattices \cite{Va06a}, 3D lattices
as well as 2D lattices with holes. On the other hand, all models
corresponding to graph states $|\varphi_{\tilde G}\rangle$ which are
not universal resources for MQC (in the sense of universal state
preparation \cite{Va06a}) are {\em not} capable of expressing
partition functions of e.g. the 2D-Ising model (or other complete
models). Examples of ``non-complete'' interaction patterns are 1D
structures such as chains or trees, or more generally all graphs
where the decorated graph $\tilde G$ has bounded rank width
\cite{Va06a}.

{\it 7.--- Summary.}
We have established a connection between evaluating the
partition function of a general class of classical spin
models and measurement-based quantum computation. We have
used the universality of the 2D cluster states, in
particular the possibility of preparing any other quantum
state by means of projective single-qubit measurements from
a sufficiently large universal state, to show a type of
completeness of the classical 2D Ising model: the partition
function of any classical spin model (Ising and Potts model on arbitrary graphs, and beyond) can be recovered as a special case
of the Ising model on a sufficiently large 2D square
lattice with complex couplings. Moreover, we have given an explicit, efficient construction of the corresponding 2D model.

Finally, it is an interesting open problem whether
a restriction to the real (and thus ``physical'') parameter
regime of the 2D-Ising model is possible while keeping the
completeness property. It would also be interesting to investigate how the explicit reductions obtained in this paper may be related to previous results regarding the NP-completeness of the 2D Ising model \cite{Ba82} (see also \cite{Sch07}).

\begin{acknowledgements}
We thank R. H\"ubener, G. Ortiz, J.I. Cirac and R. Raussendorf for interesting discussions. This work was supported by the FWF and the European Union (QICS, OLAQUI,SCALA).
\end{acknowledgements}


\begin{thebibliography}{99}
\bibitem{noteGraphstate}
Graph states are usually defined as the joint eigenstates
of the operators $K_j=X_j
\bigotimes_{k|(j,k)\in E}Z_k$ \cite{He06}. To be precise,
$|\varphi_{\tilde G}\rangle$ is equivalent to a
graph state up to local Hadamard operations on all qubits
in $V_E$.
\end{thebibliography}
\end{document}